\documentclass[aps,prl,twocolumn]{revtex4}

\usepackage{amsfonts}
\usepackage{amsmath}
\usepackage{amssymb}
\usepackage{graphicx}
\usepackage{color}
\usepackage{times}
\usepackage{float}

\newcommand{\beq}{\begin{equation}}
\newcommand{\eeq}{\end{equation}}
\newcommand{\bsp}{\begin{split}}
\newcommand{\esp}{\end{split}}
\newcommand{\bpm}{\begin{pmatrix}}
\newcommand{\epm}{\end{pmatrix}}

\newcommand{\eq}[1]{Eq. [\ref{#1}]}
\newcommand{\fig}[1]{Fig.[\ref{#1}]}

\begin{document}

\title{Localized Many-Particle Majorana Modes with Vanishing Time-Reversal Symmetry Breaking in Double Quantum Dots}

\author{Anthony R. Wright}
\email{a.wright7@uq.edu.au}
\affiliation{School of Mathematics and Physics, University of Queensland, Brisbane, 4072 Queensland, Australia}
\author{Menno Veldhorst}
\email{m.veldhorst@unsw.edu.au}
\affiliation{ARC Centre of Excellence for Quantum Computation and Communication Technology, School of Electrical Engineering \& Telecommunications, The University of New South  Wales, Sydney 2052, Australia}

\date{\today}

\begin{abstract}
We introduce the concept of spinful many-particle Majorana modes with local odd operator products, thereby preserving their local statistics. We consider a superconductor - double quantum dot system where these modes can arise  with negligible Zeeman splitting when Coulomb interactions are present. We find a reverse Mott-insulator transition, where the even and odd parity bands become degenerate. Above this transition, Majorana operators move the system between the odd parity ground state, associated with elastic co-tunneling, and the even parity ground state, associated with crossed Andreev reflection. These Majorana modes are described in terms of one, three and five operator products. Parity conservation results in a $4\pi$ periodic supercurrent in the even state and no supercurrent in the odd state. 
\end{abstract}


\maketitle
The prediction for the existence of Majorana modes \cite{Majorana1937} has attracted enormous attention in condensed matter physics \cite{Moore1991, Kitaev2003, simonrev, Fu2008, Sau2010, Alicea2010}. Allured by the possibility of constructing topological qubits for quantum computation \cite{Moore1991, simonrev}, a plethora of schemes promising the positive identification of Majorana modes has emerged \cite{fivehalves, TISC, wires, Sau2012, tony}. Although superconductors are a natural habitat of chargeless quasiparticles, spin degeneracy in standard $s$-wave superconductors prevents the appearance of localized Majorana modes. Effective spinless $p$-wave superconductors can be realized using strong spin-orbit coupling, for example at the interface of a superconductor and a topological insulator \cite{Fu2008}, or in semiconducting nanowires in the presence of Zeeman and Rashba fields \cite{Sau2010, Alicea2010}. Experimentally, supercurrents \cite{Sacepe}, Fraunhofer patterns and Shapiro steps \cite{VeldhorstJJ}, SQUIDs \cite{Veldhorstsquid}, and zero bias conductance peaks \cite{Tanaka} have been observed in topological insulator systems, which, together with the zero bias conductance peaks in nanowire systems \cite{Mourik}, provide prospects for the observation of the Majorana mode, but to date, no conclusive evidence has been observed.

\begin{figure}[t]
\centering\includegraphics[width=0.45\textwidth]{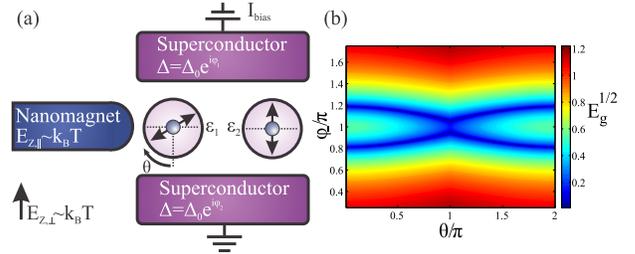}
\caption{ (a) Schematic representation of the device. Two superconductors (with phase difference $\phi_-$) are connected via a double quantum dot with on-site Coulomb repulsion $U$ in the presence of a magnetic field $E_{Z,\perp}$ that can be as small as $k_BT$. A nanomagnet introduces a localized $E_{Z,\parallel}$, which rotates the field near one of the two dots, and we define $\theta$ to be the relative angle between the local fields. (b) Tuning to the degeneracy of the ground state of the double quantum dot system. The color scale represents the energy gap between the lowest even and odd energy eigenstates, $E_g = (\epsilon_1 - \epsilon_2)/\Delta$. The parameters are $U=3\Delta=-3t$, and $E_Z = 0.2\Delta$. The degenerate ground states support many-particle Majorana modes with local statistics, for \emph{all} $\theta\ne 0$.}
\label{setupandgapplot}
\end{figure}

Superconductor - quantum dot systems are also proposed to realize Majorana modes \cite{Sau2012, Fulga, Leijnse2012}, by including spin-orbit coupling \cite{Sau2012}, or in the presence of anisotropic magnetic fields \cite{Leijnse2012}. Although quantum dots have several advantages, the strong spin-orbit or anisotropic magnetic fields required forms a major hurdle and limits material flexibility. For example, anisotropic magnetic fields only result in spinless localized Majorana modes in fields $E_Z \gg \Delta\sim t$, with $\Delta$ the induced superconducting gap and $t$ the inter-site hopping. Here, generalizing the proposal of Ref. [\onlinecite{Leijnse2012}] to strongly correlated quantum dots, we show that in the presence of small anisotropic magnetic fields, $E_Z \sim k_BT$, a new arena emerges: the concept of \emph{spinful many-particle Majorana modes}. These Majorana modes are localized in their odd operator products, and thereby preserve the local statistics of spinless proposals \cite{ivanov}. Furthermore, this demonstrates that interactions can \emph{greatly} relax the constraints of large Zeeman splitting and large spin orbit coupling, thus enhancing considerably the range of materials that may support Majorana modes. Experimentally, supercurrents through quantum dots formed in carbon nanotubes \cite{Herrero2006}, InAs nanowires \cite{Dam2006}, InAs quantum dots \cite{Buizert2007}, and graphene \cite{Dirks2007} have been observed, making them potential candidates to observe the many-particle Majorana modes. We therefore suggest that many-particle Majorana modes may become standard in practical applications of Majorana physics.     

\begin{figure}[tbp]
\centering\includegraphics[width=0.45\textwidth]{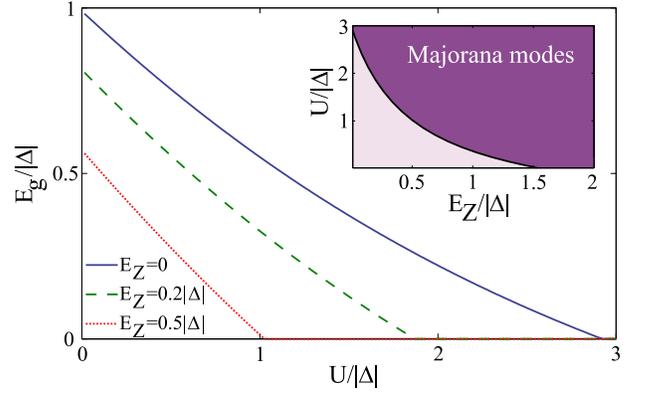}
\caption{Phase diagram showing the onset of a degenerate ground state energy gap as a function of on-site Coulomb repulsion, for $|\Delta|=|t|$ and $E_Z\rightarrow 0$. Upon increasing the on-site Coulomb repulsion $U$, a reverse Mott-insulator transition occurs where the lowest energy state changes from even to odd parity. Above this transition a degenerate ground state can be obtained using the superconducting phase $\phi_-$. The inset shows the phase diagram for finite $E_Z$. }
\label{phasediagram}
\end{figure}

We consider an $s$-wave superconductor - double quantum dot system as depicted in \fig{setupandgapplot}, where crossed Andreev reflection (CAR) dominates elastic co-tunneling (EC), a regime that is readily achieved \cite{Byers1995, Deutscher2000, Hofstetter, Hermann, menno}. Double electron occupancy in a quantum dot, as we will show, inhibits, but does not prevent the appearance of Majorana modes. On-site Coulomb repulsion $U$ can be used to tune the system from the double occupancy regime, which we define as $U=0$, towards the single electron regime, $U=\infty$. Upon increasing the Coulomb repulsion, we find a clear phase transition where it becomes possible to tune the even and odd parity states to become degenerate using a second superconductor with a phase difference $\phi_-$, see \fig{phasediagram}. Here and henceforth, parity refers to whether a state is a superposition of vectors with even or odd occupancies. Parity in this system is a good quantum number, so the even and odd sectors never mix. This is reminiscent of a Mott metal-insulator transition, however in reverse, where the on-site Coulomb repulsion drives the system \emph{toward} having a degenerate ground state.

In order to clearly elucidate the novel many-particle physics that emerges, we will present most results in the limit of negligibly small Zeeman splitting, $E_Z$, valid as $T\rightarrow 0$, meaning we have spinful, nearly spin rotation symmetric, ground states. However, we emphasize that $|E_Z| > 0$ is a strict necessary condition to obtain a unique odd-parity ground state, by Kramer's degeneracy.

The superconductor - double quantum dot system with on-site Coulomb repulsion in the presence of an anisotropic magnetic field, \fig{setupandgapplot}, is described by the Hamiltionian
\beq
H = H_S + H_D + H_T + H_Z.
\label{H_all}
\eeq
The superconducting part, $H_S$, is two regular $s$-wave superconducting leads, given by
\beq
H_S = \sum_{i=1}^2\sum_{k} \biggl(\sum_\sigma\xi_{k,i,\sigma} d_{k,i,\sigma}^\dag d_{k,i,\sigma}\biggr) + \Delta e^{i\phi_i}d_{k,i,\uparrow}^\dag d_{-k,i,\bar\downarrow}^\dag + h.c.,
\eeq
where $d_{k,i,\sigma}^\dag$ creates an electron in the $i^\mathrm{th}$ superconductor \cite{Eldridge2010} with momentum $k$ and spin $\sigma$, $\xi_{k,i,\sigma}$ is the non-interacting dispersion, $\Delta$ is the superconducting amplitude (assumed equal in the two leads for simplicity) and $\phi_i$ the superconducting phase. Here, $\bar\sigma$ denotes the spin that is not $\sigma$. The double quantum dot is described by, $H_D$,
\beq
H_D = \sum_{j=1}^2\sum_{\sigma}\epsilon_jc_{j,\sigma}^\dag c_{j,\sigma} + \sum_j Un_{j,\uparrow}n_{j,\downarrow},
\eeq
with $c_j$ creating an electron in the $j^\mathrm{th}$ dot, $U$ the onsite Coulomb repulsion, $n$ the number operator, and $\epsilon_j$ the onsite energy. The third term in \eq{H_all} introduces the tunneling between the dots and superconductors and is described by $H_T$,
\beq
H_T = \sum_i^2\sum_j^2\sum_\sigma\Gamma_{ij} d_{i,\sigma}^\dag c_{j,\sigma} + h.c.
\eeq
The overlap integral $\Gamma_{ij}$ is between the end of the $i^{\mathrm{th}}$ superconducting lead, and the $j^\mathrm{th}$ dot, and is assumed to be equal in all cases (i.e., $\Gamma_{ij} = \Gamma$). Finally, the magnetic field $E_Z$ results in

\beq
H_Z = -E_Z\sum_{j}\bigl(c_{j,\uparrow}^\dag c_{j,\uparrow} - c_{j,\downarrow}^\dag c_{j,\downarrow}\bigr).
\eeq

The anisotropic magnetic field provides the most convenient definition of the spin axes. The angle $\theta$, the angle between the two local magnetic fields, modify the EC and CAR as follows:
\begin{equation}
\begin{array}{rl}
tc_{1,\sigma}^\dag cs_{2,\sigma}\rightarrow& t\cos(\theta/2)c_{1,\sigma}^\dag c_{2,\sigma} + \sigma t\sin(\theta/2)c_{1,\sigma}^\dag c_{2,\bar\sigma}\\
\Delta c_{1,\sigma}c_{2,\bar\sigma}\rightarrow& \bar\sigma\Delta\sin(\theta/2)c_{1,\sigma}c_{2,\sigma} + \Delta\cos(\theta/2)c_{1,\sigma}c_{2,\bar\sigma},
\end{array} 
\label{angles}
\end{equation}
where $\sigma,\bar\sigma = \pm$, and $t,\Delta$ are the effective EC hopping and CAR Cooper pairing amplitudes \cite{arovas}. 

The Hamiltonian is quadratic in the leads, and so we can integrate them out, following the procedure of Ref.[\onlinecite{arovas}], to obtain the effective Hamiltonian
\beq
\bsp
H_{eff} = &\sum_{j,\sigma} \epsilon_j c_{j,\sigma}^\dag c_{j,\sigma} + U\sum_j n_{j,\uparrow}n_{j,\downarrow} - E_Z\sum_j\bigl(n_{j,\uparrow} - n_{j,\downarrow}\bigr)\\
+& \Bigl[\sum_\sigma \Bigl(t\cos(\theta/2)c^\dag_{1,\sigma}c_{2,\sigma} + \sigma t\sin(\theta/2)c^\dag_{1,\sigma}c_{2,\bar\sigma}\\
+& \sigma\Delta e^{i\frac{\phi_+}{2}}\cos\bigl(\frac{\phi_-}{2}\bigr)\bigl(\bar\sigma \sin(\theta/2)c^\dag_{1,\sigma}c^\dag_{2,\sigma} + \cos(\theta/2)c^\dag_{1,\sigma}c^\dag_{2,\bar\sigma}\bigr) \Bigr)\\
+& \Delta e^{i\frac{\phi_+}{2}}\cos\bigl(\frac{\phi_-}{2}\bigr)\sum_j c^\dag_{j,\uparrow}c^\dag_{j,\downarrow} + h.c.\Bigr],
\end{split}
\label{Heff}
\eeq
where $\phi_\pm = \phi_1 \pm \phi_2$ is the sum (difference) between the phases of the two superconductors. From hereon we will assume that the onsite energy of the two dots has been tuned to the chemical potential of the superconductors, which we define as our zero of energy ($\epsilon_1=\epsilon_2=0$). A discussion of the effects of the onsite energies deviating from this `sweet spot' has been presented elsewhere \cite{Leijnse2012}. 

The Hamiltonian \eq{Heff} cannot be decomposed into total spin sectors, as the anisotropic magnetic field mixes these. However, the mode parity is a conserved quantity. The total Hilbert space has dimension $(2^{n_{\sigma}})^{n_{j}}/2 = 8$ for each parity sector (i.e. even and odd), which makes exact diagonalization particularly straight-forward. In the occupation representation, we can define an occupation basis $|1, 2\rangle$, where the numbers correspond to the two dots, and we will use arrows to denote the spin, and then we can construct a basis of the sixteen possible configurations.  In the $U\rightarrow \infty$ limit, the total Fock space is restricted to nine possible states, $5$ for even parity, and $4$ for the odd. We solve for finite Zeeman splitting, and obtain in the odd parity sectors that the eigenvalues for the lowest energy odd state is $\epsilon_{odd} = - t + E_Z\cos(\theta/2),\,\,\,(t>0)$, and for the even state, $\epsilon_{even} = -\sqrt{2}\Delta \cos(\phi_-/2)$. When $T \rightarrow 0$, the required Zeeman splitting becomes negligibly small relative to $\Delta$. The crucial role of the Zeeman field is to break Kramer’s degeneracy, and thus to define the lowest energy odd parity eigenstate. Having determined this state, we consider it in the limit of zero field. The corresponding wavefunctions are then $ \Psi_{even} = \frac{1}{2}\Bigl[\sqrt{2}e^{\frac{i\phi_+}{2}}|0,0\rangle+\cos(\theta/2)\bigl(|\uparrow,\downarrow\rangle - |\downarrow,\uparrow\rangle\bigr)+\sin(\theta/2)\bigl(|\uparrow,\uparrow\rangle + |\downarrow,\downarrow\rangle\bigr)\Bigr]$, and $\Psi_{odd} = \frac{1}{\sqrt{2}}\Bigl[\sin(\theta/4) \\  \bigl(|\uparrow, 0\rangle+ |0,\uparrow\rangle\bigr)+ \cos(\theta/4)\bigl(|\downarrow,0\rangle - |0,\downarrow\rangle\bigr)\Bigr]$. A degenerate ground state is obtained when $\epsilon_{even} = \epsilon_{odd}$. Crucially, for any $t+E_Z<\sqrt{2}|\Delta|$, there is always a $\phi_-$ which can be chosen, such that a degenerate ground state can be obtained. 

We emphasize that as we are using the occupation number basis, \emph{the ground state is given by the lowest energy eigenstate}, and not the zero energy eigenstate as in the more familiar Bogoliubov de-Gennes theory.

The two degenerate ground states are protected from hybridizing when the total system conserves particle number parity, and a pair of Majorana operators ($\gamma_1,\gamma_2$) can be constructed which transform the two ground states into each other, such that $\gamma_{1}\Psi_{odd} = \Psi_{even}$, for example. These Majorana operators are given by
\beq
\bsp
\gamma^{U\rightarrow\infty}_1&(\theta) = \frac{1}{\sqrt{2}}\Biggl[x(\gamma_{1\uparrow}(1-n_{1\downarrow}) + \gamma_{2\downarrow}(1-n_{1\uparrow}))\\
&- \Bigl(\frac{\cos(\theta/2)}{\cos(\theta/4)} + x\Bigr)n_{2\downarrow}\gamma_{1\uparrow} +\Bigl(\frac{\sin(\theta/2)}{\sin(\theta/4)} - x\Bigr)n_{2\uparrow}\gamma_{1\uparrow}\\
& - \Bigl(\frac{\cos(\theta/2)}{\sin(\theta/4)} + x\Bigr)n_{2\uparrow}\gamma_{1\downarrow}-\Bigl(\frac{\sin(\theta/2)}{\cos(\theta/4)} + x\Bigr)n_{2\downarrow}\gamma_{1\downarrow}\biggr]
\end{split}
\label{gam1}
\eeq    
where $x = (\sin(\theta/4) + \cos(\theta/4))^{-1} $, and $\gamma_{1,\uparrow}^{\phi_+} = (e^{i\phi_+/4}c_{1,\uparrow}^\dag + e^{-i\phi_+/4}c_{1,\uparrow})$, and $n_{i,\sigma} = c^\dag_{i,\sigma}c_{i,\sigma}$. $\gamma^{U\rightarrow\infty}_2(\theta)$ has a similar form with site indices $1,2$ interchanged. $\gamma_{1,\uparrow}^{\phi_+}$ has the form of a usual Majorana operator \cite{ivanov}, except that the phase dependence is the total phase of the two superconductors \emph{divided by four}, or equivalently half the average phase of the two superconductors. It is assumed, in the above Majorana expressions, that the system is projected into the singly occupied Fock space, since $U\rightarrow\infty$.

When $U$ is finite, a phase transition occurs at a critical value of the on-site repulsion. For Coulomb repulsions below this critical point, the ground state is non-degenerate, with the lowest energy even parity state always having lower energy than the lowest energy odd parity state. In \fig{phasediagram} we have plotted the excitation energy of the first excited state $E_g = (\epsilon_2-\epsilon_1)/|\Delta|$ as a function of onsite Coulomb repulsion. The phase transition corresponds to the vanishing of the excitation gap at a critical value of on-site Coulomb repulsion, of the same order as the effective pairing $|\Delta|$.

For finite on-site Coulomb repulsion, the odd parity ground state develops a finite weighting on the terms of the form $|\uparrow\downarrow,\uparrow\rangle$, together with the three equivalent combinations of this, whilst the even parity ground state develops a finite weighting on doubly occupied dots ($|\uparrow\downarrow,0\rangle$ and $|0,\uparrow\downarrow\rangle$), together with the four mode, double occupied quantum dot pair $|\uparrow\downarrow,\uparrow\downarrow\rangle$. The Majorana modes in this case acquire five mode operator products, together with single and three mode operator products. Generically, the Majorana mode has the form

\beq
\gamma_1 = \gamma_{1,\uparrow}^{\phi_+}\bigl(a^\uparrow + \sum_{\sigma j} b^\uparrow_{\sigma j} n_{\sigma j} + \sum_{\sigma \sigma' j j'} c^\uparrow_{\sigma \sigma' j j'} n_{\sigma j}n_{\sigma' j'}\bigr) + \uparrow\leftrightarrow\downarrow,
\label{5operator}
\eeq
where $b^\sigma_{\sigma,1} = c^{\sigma}_{\sigma,\sigma',1,j'} = c^{\sigma}_{\sigma',\sigma,j,1} = 0$, and when $\phi = \pi/2$, all operators where $j,j'=2$ have coefficient zero. The coefficients $a,b$ and $c$, which determine the relative weights of the even operator products, (i.e., $n_{\sigma,i} = c^\dag_{\sigma,i}c_{\sigma,i}$), depend on $U$ and $\theta$. However, the odd operator products do not and the Majorana modes are therefore robust against variations in the Coulomb and magnetic fields. The nonlocality of the Majorana modes in \eq{gam1} and \eq{5operator} is restricted to number operators only. Therefore, the relative phase of the Majorana components, which is responsible in general for their non-Abelian braiding statistics, is spatially isolated and localized on a single dot. The appearance of three and five operator products are a clear generalization of the Majorana mode concept, which is usually based on single operator products \cite{nayak, ivanov}.

\begin{figure}[tbp]
\centering\includegraphics[width=0.45\textwidth]{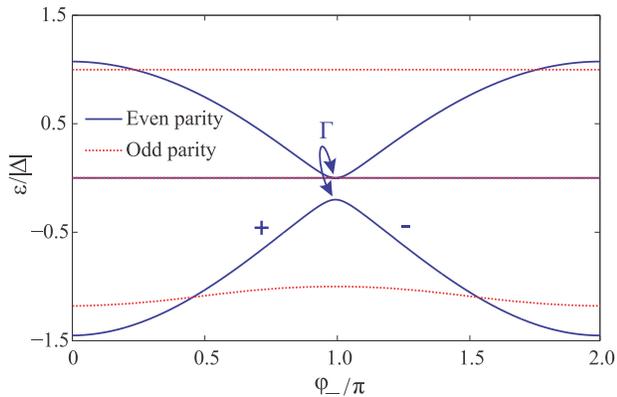}
\caption{Andreev bound states of the double dot system for $U$=20$|\Delta|$, $|\Delta|=|t|$, $\theta=\pi/2$ and $E_Z \rightarrow 0$. The non-dispersive states at $E=0$ correspond to trivial even and odd solutions. The $+,-$ signs indicate the even parity ground state bonding and antibonding state, determined by the relative sign between the empty and two-mode occupations, for example $\Psi_\pm = a|0000\rangle \pm b|1010\rangle + ...$. At infinite $U$, the two states are orthogonal, but at finite $U$ the two develop an anticrossing, thereby opening a gap which increases with decreasing $U$. By means of Zener tunneling $\Gamma$, quasiparticles can overcome small hybridization gaps, so that a 4 $\pi$ periodic Josephson effect can be observed at finite $U$.}
\label{BAB}
\end{figure}

The regime where Majorana modes appear can be found by analyzing the Josepshon supercurrent through the double dot system. The Josephson current is calculated from the derivative of the free energy with respect to the superconducting phase difference $\phi_-$ \cite{droste}. In \fig{BAB} we show the Andreev bound states for finite Coulomb interaction $U=20\Delta$. Whereas in non-degenerate systems such as topological superconductors, the even and odd parity ground states disperse equally and oppositely \cite{Kitaev2003}, the degeneracy in dots leads to odd parity ground states that disperse only weakly with the superconducting phase difference, and are strictly $2\pi$ periodic. At $U\rightarrow\infty$, the odd parity ground states are completely flat, as CAR cannot possibly excite these states, while the even parity ground states are gapless and 4$\pi$ periodic. At finite $U$, the bonding and antibonding even parity modes hybridize via the fully occupied $|\uparrow\downarrow,\uparrow\downarrow\rangle$ state and a gap opens, and the odd parity states disperse weakly via CAR into the triply occupied states. The even state is 4$\pi$ periodic for $U=\infty$, but by means of Zener tunneling $\Gamma$, quasiparticles can tunnel from a lower to a higher Andreev bound state \cite{Jacobs, Kroemer}, and a 4$\pi$ periodic contribution is expected for a large range of finite $U$. This could be measured in voltage biased experiments, for example by measuring Shapiro steps. This transition from $4\pi$ to 2$\pi$ periodicity is not a transition where Majorana modes disappear, but a transition where the five operator products, \eq{5operator}, appear.

\begin{figure}[tbp]
\centering\includegraphics[width=0.45\textwidth]{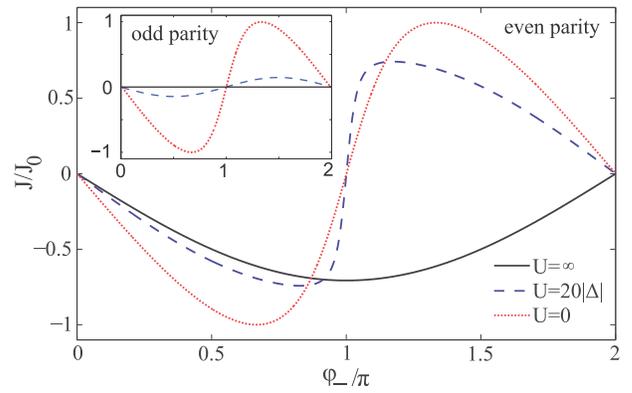}
\caption{Josephson supercurrent at $T=0.01|\Delta|$ for Coulomb interaction $U=0,20|\Delta|$ and $\infty$, with $|\Delta|=|t|$ and $\theta=\pi/2$, with $J_0=e\Delta/\hbar$, the maximum supercurrent at $U=0$. When $U=0$, local Andreev reflection dominates, no Majorana modes exist and the supercurrent for the even and odd parity sectors are equal. When $U \rightarrow \infty$, the Andreev reflection is purely non-local, the supercurrent in the even parity is $4\pi$ periodic, and there is no supercurrent in the odd parity state. For finite $U$, hybridization between the even parity eigenstates leads to a sharp transition at $\phi=\pi$ in the even parity state, and doubly occupied modes support a small supercurrent in the odd parity state. Majorana modes are present for all $U$ above the Mott-insulator transition.}
\label{Currents}
\end{figure}

In \fig{Currents} we have plotted the current-phase relation for different Coulomb interactions. When $U\rightarrow \infty$ the Cooper pairs split solely via CAR and there is a 4$\pi$ periodic even parity supercurrent, while there is no supercurrent in the odd parity state. \emph{The absence of supercurrent in the odd parity state will be a strong signature of parity conservation.} When $U=0$, AR dominates and the even and odd states carry equal supercurrent. Increasing $U$ results in a sharper kink of the supercurrent in the even parity sector around $\phi=\pi$ due to a decreasing hybridization gap, and a gradual vanishing of the odd parity state supercurrent.  Contrary to the non-degenerate $p$-wave superconductors, the strongly dispersive bound states in the double quantum dot system both have even parity, and the branches are not parity protected. The anomalous current phase relationship, however, can still be observed in non-equilibrium measurements \cite{Badiane2011}, or using DC SQUID's \cite{Veldhorst20121, Veldhorst20122, Choi2000}.

In conclusion, we have introduced the concept of spinful many-particle Majorana modes, and showed that these can be realized in double quantum dots. Coulomb interactions greatly relax the constraints of large Zeeman splitting and spin-orbit coupling. When two superconductors are connected to both quantum dots, the superconducting phase can be used to obtain a degenerate ground state protected by parity, with Majorana modes constructed from one, three and five mode creation/annihilation operator products connecting the ground states. The relative amplitudes of the operators of the many-particle Majorana modes are dependent on the field angle  $\theta$  and the Coulomb interaction $U$. However, these weightings do not affect the odd operator products, which are localized on a dot and are responsible for their non-Abelian braiding statistics. We expect therefore that the arising many-particle Majorana modes will be as robust as the Majorana modes in spinless proposals. The effect of a finite, rather than infinite, on-site interaction does not affect the locality conditions of the Majoranas, nor their parity. We expect that the concept of many-particle Majorana modes can also be realized in topological superconductors, where Coulomb interactions might relax the constraints in those systems as well.

\acknowledgments
We would like to thank Ross H. McKenzie, Thomas E. O'Brien, Jacopo Sabatini, Marieke Snelder, and Andrew Dzurak  for enlightening discussions. We also thank Michele Governale, Stephanie Droste, Martin Leijnse and Karsten Flensberg for comprehensive manuscript feedback. ARW is financially supported by a University of Queensland Postdoctoral Reasearch Fellowship. MV is financially supported by the Australian Research Council Centre of Excellence for Quantum Computation and Communication Technology (project number CE11E0096), the US Army Research Office (W911NF-13-1-0024) and the Netherlands Organization for Scientific Research (NWO) by a Rubicon grant.

\end{document}